\documentclass[12pt]{amsart}

\usepackage{amsmath,amsthm,amsfonts,amscd,amssymb}
\usepackage{times,euler,eucal,eufrak,graphicx}
\usepackage{pst-node}
\usepackage{pst-plot}

\usepackage{hyperref}

\title[Integration on the unitary group]
{Integration with respect to the Haar measure on unitary,
orthogonal and symplectic group}
\thanks{B.C. is supported by a JSPS postdoctoral fellowship}
\author {Beno\^\i{}t Collins}
\address{Department of Mathematics, Graduate School of Science,
Kyoto university, Kyoto 606-8502, Japan} \email{collins@math.kyoto-u.ac.jp}

\author{Piotr \'Sniady}
\thanks{P.\'S.~was supported by State Committee for Scientific Research (KBN)
grant \mbox{2 P03A 007 23}}
\address{Institute of Mathematics,
University of Wroclaw, pl.\ Grunwaldzki 2/4, 50-384 Wroclaw,
Poland} \email{Piotr.Sniady@math.uni.wroc.pl}

\theoremstyle{plain}
\newtheorem{lemma}{Lemma}[section]
\newtheorem{theorem}[lemma]{Theorem}
\newtheorem{proposition}[lemma]{Proposition}
\newtheorem{corollary}[lemma]{Corollary}

\theoremstyle{definition}

\theoremstyle{remark}
\newtheorem*{remark}{Remark}
\newtheorem*{example}{Example}

\DeclareMathOperator{\tr}{tr} \DeclareMathOperator{\Tr}{Tr}
\DeclareMathOperator{\We}{Wg} \DeclareMathOperator{\Wg}{Wg}
\DeclareMathOperator{\Id}{Id} \DeclareMathOperator{\End}{End}
\DeclareMathOperator{\Vect}{Vect}
\DeclareMathOperator{\Moeb}{Moeb}
 \DeclareMathOperator{\GL}{GL}
 \DeclareMathOperator{\U}{U}
\DeclareMathOperator{\Ort}{O} %\DeclareMathOperator{\Sp}{Sp}
\DeclareMathOperator{\Smp}{Sp} \DeclareMathOperator{\B}{B}

\newcommand{\M}[1]{M_{#1}(\mathbb{C})}
\newcommand{\Sy}[1]{\mathcal{S}_{#1}}
\newcommand{\E}{{\mathbb{E}}}
\newcommand{\A}{{\mathcal{A}}}
\newcommand{\C}{{\mathbb{C}}}

\newcommand{\N}{{\mathbb{N}}}
\newcommand{\gwia}{^{\star}}
\newcommand{\reps}{\rho_{\Sy{n}}^{d}}
\newcommand{\repu}{\rho_{\U(d)}^{n}}

\newcommand{\repo}{\rho_{\Ort(d)}^{n}}

\newcommand{\dist}{l}

\newcommand{\eigen}{z_{\lambda}}

\begin{document}

\begin{abstract}
We revisit the work of the first named author and using simpler
algebraic arguments we calculate integrals of polynomial functions
with respect to the Haar measure on the unitary group $\U(d)$. The
previous result provided exact formulas only for $2d$ bigger than
the degree of the integrated polynomial and we show that these
formulas remain valid for all values of $d$. Also, we consider the
integrals of polynomial functions on the orthogonal group
$\Ort(d)$ and the symplectic group $\Smp(d)$. We obtain an exact
character expansion and the asymptotic
behavior for large $d$. Thus we can show the asymptotic freeness
of Haar-distributed orthogonal and symplectic random matrices, as
well as the convergence of integrals of the Itzykson--Zuber type.
\end{abstract}

\maketitle

%\tableofcontents

%%%%%%%%%%%%%%%%%%%%%%%%%%%%%%%%%%%%%%%%%%
%%%%%%%%   NOTATION
%%%%%%%%
%%%%%%%%   \q            number of factors
%%%%%%%%   \dd           dimension of C^d
%%%%%%%%   \dist         length on Brauer algebra

\section{Introduction}
\label{sec:intro}

Let $G\subset \End(\C^d)$ be a compact Lie group viewed as a group
of matrices. The matrix structure provides a very natural
coordinate system on $G$; in particular we are interested in the
family of functions $e_{ij}:G\rightarrow\C$ defined by $e_{ij}:
\M{d} \ni m \mapsto m_{ij}$ which to a matrix  assign one of its
entries. We call polynomials in $(e_{ij})$ polynomial functions on
$G$. In this article we are interested in the integrals of
polynomial functions in $(e_{ij},\overline{e_{ij}})$ on compact
Lie groups with respect to the Haar measure on $G$, i.e. the
integrals of the form
\begin{equation}
\int_G  U_{i_1 j_1} \cdots U_{i_n j_n} \overline{U_{i_1' j_1'}}
\cdots \overline{U_{i_{n'}' j_{n'}'}} \ dU. \label{eq:calka}
\end{equation}
For simplicity, such integrals will be called moments of the group
$G$.

If we consider a matrix-valued random variable $U$ the
distribution of which is the Haar measure on $G$ then the
integrals of the form \eqref{eq:calka} have a natural
interpretation as certain moments of entries of $U$ and they
appear very naturally in the random matrix theory. The reason for
this is that quite many random matrix ensembles $X$ are invariant
with respect to the conjugation by elements of the group $G$ and
therefore can be written as $X=U X' U^{-1}$, where $U$ and $X'$
are independent matrix-valued random variables and the
distribution of $U$ is the Haar measure on $G$. As a result, the
expressions similar to
\begin{equation}
\E \Tr (X_1 U^{s_1} X_2 U^{s_2} \cdots X_n U^{s_n})
\label{eq:przyklad}
\end{equation}
are quite common in the random matrix theory, where
$s_1,\dots,s_n\in\{1,\star\}$ and $X_1,\dots,X_n$ are some
matrix-valued random matrices independent from $U$. It is easy to
see that the calculation of \eqref{eq:przyklad} can be easily
reduced to the calculation of \eqref{eq:calka}. In the random
matrix theory we are quite often interested not in the exact value
of the expression of type \eqref{eq:przyklad} but in its
asymptotic behavior if $d$ tends to infinity. The results of this
type were obtained for the first time by Weingarten
\cite{Weingarten1976}.

In this article we are interested in the case when $G\subset
\M{d}$ belongs to one of the series of the classical Lie groups,
i.e.~$G$ is either the unitary group $\U(d)$ or the orthogonal
group $\Ort(d)$ or the symplectic group $\Smp(d/2)$, where in the
latter case we assume that $d$ is even. Firstly, we revisit a part
of the work of the first named author \cite{Collins2002} and
compute with a new convolution formula the moments of the unitary
group. This formula gives a new combinatorial insight into the
relation between free probability and asymptotics of moments of
the unitary group. Then, we make use of other features of
invariant theory to give an explicit integration formula on the
orthogonal and symplectic groups and to compute asymptotics in the
latter case. This allows us to prove a new convergence result for
a large family of matrix integrals. Our main tool is the
Schur--Weyl duality for the unitary group and its analogues for
the orthogonal and symplectic groups.

\section{Integration over unitary groups}

\subsection{Schur--Weyl duality for unitary groups}

We recall a couple of notations and standard facts. A
non-increasing sequence of nonnegative integers $\lambda
=(\lambda_1,\ldots )$ is said to be a partition of the integer $n$
(abbreviated by $\lambda\vdash n$) if $\sum_i\lambda_i=n$. We
denote by $l(\lambda )$ its length, i.e. the largest index $i$ for
which $\lambda_i$ is non-zero.

There is a canonical way to parameterize all irreducible
polynomial representations
$\rho^{\lambda}_{\U(d)}:\U(d)\rightarrow \End V^{\lambda}_{\U(d)}$
of the compact unitary group $\U(d)$ by partitions $\lambda$ such
that $l(\lambda)\leq d$. The character of this representation
evaluated on the torus is the Schur polynomial $s_{\lambda,d}$
(see \cite{FultonYoungtableaux}). By $s_{\lambda,d}(x)$ we shall
understand $s_{\lambda,d}(x,\ldots ,x)$ with $d$ copies of $x$. In
particular, $s_{\lambda,d}(1)$ is the dimension of the
representation $V^{\lambda}_{\U(d)}$ of $\U(d)$.

The group algebra $\C[\Sy{n}]$ of the symmetric group $\Sy{n}$ is
semi-simple. It is endowed with its canonical basis
$\{\delta_{\sigma}\}_{\sigma\in\Sy{n}}$. The irreducible
representations $\rho^{\lambda}_{\Sy{n}}:\Sy{n}\rightarrow \End
V^{\lambda}_{\Sy{n}}$ are canonically labelled by $\lambda\vdash
n$ via the Schur functor (see \cite{FultonYoungtableaux} as well);
we denote the corresponding characters by $\chi^\lambda$.

The following isomorphism holds:
\begin{equation}
\label{eq:decompositionSq} \C[\Sy{n}] \cong
\bigoplus_{\lambda\vdash n}   \End V^{\lambda}_{\Sy{n}}.
\end{equation}
For any $\lambda\vdash n$, let $p^{\lambda}=
\frac{\chi^{\lambda}(e)}{n!} \chi^{\lambda}\in\C[\Sy{n}]$ be the
minimal central projector onto $\End V^{\lambda}_{\Sy{n}}$. We
define for future use the algebra
\begin{equation}
\label{eq:definitionofcd} \C_d[\Sy{n}]=\bigg(\sum_{\lambda\vdash
n,\ l(\lambda)\leq d}p^{\lambda} \bigg) \C[\Sy{n}]=
\bigoplus_{{\lambda\vdash n,\ l(\lambda)\leq d}} \End
V^{\lambda}_{\Sy{n}}.
\end{equation}

Consider the representation $\reps$ of $\Sy{n}$ on
$(\C^d)^{\otimes n}$, where
$$\reps(\pi):v_1\otimes\cdots\otimes v_n\mapsto
    v_{\pi^{-1}(1)}\otimes\cdots\otimes v_{\pi^{-1}(n)}$$
is given by natural permutation of elementary tensors. We consider
also the representation $\repu$ of $\U(d)$ on $(\C^d)^{\otimes
n}$, where
$$\repu(U):v_1\otimes\cdots\otimes v_n\mapsto
   U( v_{1})\otimes\cdots\otimes U(v_{n})$$
is the diagonal action. Since the representations $\reps$ and
$\repu$ commute, we obtain a representation $\rho_{\Sy{n}\times
\U(d)}$ of $\Sy{n}\times \U(d)$ on $(\C^d)^{\otimes n}$.

\begin{theorem}[Schur--Weyl duality for unitary groups \cite{WeylClassicalGroups}]
\label{theo:swunitary} The action of  $\Sy{n}\times\U(d)$ is
multiplicity free, i.e.~no irreducible representation  of
$\Sy{n}\times \U(d)$ occurs more than once in $\rho_{\Sy{n}\times
{\U(d)}}$. The decomposition of $\rho_{\Sy{n}\times {\U(d)}}$ into
irreducible components is given by
\begin{equation}
(\C^d)^{\otimes n} \cong \bigoplus_{\lambda\vdash n,\
l(\lambda)\leq d} V^{\lambda}_{\Sy{n}} \otimes
V^{\lambda}_{\U(d)}, \label{eq:decomposition}
\end{equation}
where $\Sy{n}\times \U(d)$ acts by $\rho^{\lambda}_{\Sy{n}}\otimes
\rho^{\lambda}_{\U(d)}$ on the summand corresponding to $\lambda$.
\end{theorem}

We shall consider the inclusion of algebras
\begin{equation*}
\reps (\C_d[\Sy{n}]) \subseteq\End (\C^d)^{\otimes n}.
\end{equation*}
Equations \eqref{eq:definitionofcd} and \eqref{eq:decomposition}
show that $\reps$ is injective when restricted to $\C_d[\Sy{n}]$
and for this reason we shall omit $\reps$ whenever convenient and
consider $\C_d[\Sy{n}]$ as sitting inside $\End (\C^d)^{\otimes
n}$. Conversely, we can identify every element of the image $\reps
(\C_d[\Sy{n}])\subseteq \End (\C^{d})^{\otimes n}$ with the unique
corresponding element of the group algebra $\C_d[\Sy{n}]$.

\subsection{Conditional expectation}
For $A\in\End (\C^d)^{\otimes n}$ we define
\begin{equation}
\label{eq:conditional} \E (A)=\int_{\U(d)} U^{\otimes n}\ A\
(U^{-1})^{\otimes n}\ dU,
\end{equation}
where the integration is taken with respect to the Haar measure on
the compact group $\U(d)$.

We recall that for an algebra inclusion $M\subset N$, a
conditional expectation is a $M$-bimodule map $\E : N\rightarrow
M$ such that $\E(1_N)=1_M$.

\begin{proposition}
\label{prop:conditional} $\E$ defined in \eqref{eq:conditional} is
a conditional expectation of \/ $\End (\C^d)^{\otimes n}$ onto
$\C_d[\Sy{n}]$. We regard $\End  (\C^d)^{\otimes n}$ as an
Euclidean space with a scalar product $\langle A, B\rangle= \Tr
A^\ast B$. Then $\E$ is an orthogonal projection onto $\reps\big(
\C_d[\Sy{n}] \big).$ Moreover, it is compatible with the trace in
the sense that
\begin{equation*}
\Tr\circ \E =\Tr.
\end{equation*}
\end{proposition}
\begin{proof}
Since Haar measure is a probability measure invariant with respect
to the left and right multiplication therefore $\E(A)$ commutes
with the action of the unitary group $\U(d)$ for every $A\in\End
(\C^d)^{\otimes n}$. Theorem \ref{theo:swunitary} shows that
$\E(A)\in \C_d[\Sy{n}]$ and that the range of $\E$ is exactly
$\C_d[\Sy{n}]$. Since $\langle \E(A),\E(B) \rangle= \langle \E(A),
B \rangle$ it follows that $\E$ is an orthogonal projection. The
other statements of the Proposition can be easily checked
directly.
\end{proof}

For $A\in \End (\C^d)^{\otimes n}$ we set
\begin{equation}
\Phi (A) = \sum_{\sigma\in \Sy{n}}\Tr \big(A\ \reps(\sigma^{-1})
\big)\delta_{\sigma} \in \C(\Sy{n}).
\end{equation}

\begin{proposition}
\label{prop:propertiesofPhi} $\Phi$ fulfils the following
properties:

\renewcommand{\labelenumi}{\arabic{enumi}.}

\begin{enumerate}

\item \label{Phi:point1} $\Phi$ is a $\C[\Sy{n}]$--$\C[\Sy{n}]$
bimodule morphism in the sense that
\begin{align*}
\Phi \big(A\ \reps (\sigma)\big)= & \Phi (A)\ \sigma,\\
\Phi \big(\reps(\sigma)\ A\big)= & \sigma\ \Phi (A);
\end{align*}

\item \label{Phi:point2} $\Phi(\Id)$ coincides with the character
of $\reps$ hence it is equal to
\begin{equation}
\label{eq:formulanaphiid} \Phi(\Id) = n! \sum_{\lambda\vdash n}
\frac{ s_{\lambda ,d}(1)}{\chi^{\lambda}(e)} \ p^{\lambda}
\end{equation}
and is an invertible element of $\C_d[\Sy{n}]$; its inverse will
be called Weingarten function and is equal to
\begin{equation}
\label{eq:wg} \Wg=\frac{1}{(n!)^2} \sum_{\substack{\lambda \vdash n \\
l(\lambda)\leq d}} \frac{\chi^{\lambda}(e)^2}{s_{\lambda ,d}(1)}\
\chi^{\lambda}
\end{equation}

\item \label{Phi:point4} the relation between $\Phi(A)$ and
$\E(A)$ is explicitly given by
$$\Phi (A)=\E(A)\Phi (\Id);$$
\item \label{Phi:point3} the range of $\Phi$ is equal to
$\C_d[\Sy{n}]$;

\item \label{numero1} in $\C_d[\Sy{n}]$, the following holds true
\begin{equation}\label{eq:glowne}
\Phi (A\ \E (B))=\Phi (A)\Phi (B)\Phi (\Id)^{-1}.
\end{equation}

\end{enumerate}
\end{proposition}

\begin{proof}
Points \ref{Phi:point1} and \ref{Phi:point2} are immediate. Point
\ref{Phi:point1} implies
$$\Phi(A)=\Phi\big( \E(A) \big)=\Phi\big( \Id \ \E(A) \big)= \Phi(\Id)\ \E(A)$$
which proves point \ref{Phi:point4}. Point \ref{Phi:point3}
follows from point \ref{Phi:point4} and point \ref{Phi:point2}.
Point \ref{numero1} follows from points \ref{Phi:point1} and
\ref{Phi:point4}.
\end{proof}

\begin{corollary}
\label{cor:calka} Let $n$ be a positive integer and
$\mathbf{i}=(i_1,\ldots ,i_n)$, $\mathbf{i'}=(i'_1,\ldots
,i'_{n})$, $\mathbf{j}=(j_1,\ldots ,j_n)$,
$\mathbf{j'}=(j'_1,\ldots ,j'_{n})$ be $n$-tuples of positive
integers. Then
\begin{multline}
\label{bid} \int_{\U(d)}U_{i_{1}j_{1}} \cdots U_{i_{n}j_{n}}
\overline{U_{i'_{1}j'_{1}}} \cdots
\overline{U_{i'_{n}j'_{n}}}\ dU=\\
\sum_{\sigma, \tau\in\Sy{n}}\delta_{i_1i'_{\sigma (1)}}\ldots
\delta_{i_n i'_{\sigma (n)}}\delta_{j_1j'_{\tau (1)}}\ldots
\delta_{j_n j'_{\tau (n)}}\Wg (\tau\sigma^{-1}).
\end{multline}

If $n\neq n'$ then
\begin{equation} \label{bidula} \int_{\U(d)}U_{i_{1}j_{1}} \cdots
U_{i_{n}j_{n}} \overline{U_{i'_{1}j'_{1}}} \cdots
\overline{U_{i'_{n'}j'_{n'}}}\ dU= 0.
\end{equation}
\end{corollary}

\begin{proof}
In order to show \eqref{bid} it is enough to take appropriate $A$
and $B$ in $\M{d}^{\otimes n}$ and take the value of both sides of
\eqref{eq:glowne} in $e\in\Sy{n}$.

For every $u\in\C$ such that $|u|=1$ the map $\U(d)\ni U\mapsto u
U \in \U(d)$ is measure preserving therefore \begin{multline*}
\int_{\U(d)}U_{i_{1}j_{1}} \cdots U_{i_{n}j_{n}}
\overline{U_{i'_{1}j'_{1}}} \cdots \overline{U_{i'_{n'}j'_{n'}}}\
dU=\\  \int_{\U(d)} u U_{i_{1}j_{1}} \cdots u U_{i_{n}j_{n}}
\overline{u U_{i'_{1}j'_{1}}} \cdots \overline{u
U_{i'_{n'}j'_{n'}}}\ dU
\end{multline*}
and \eqref{bidula} follows.
\end{proof}

The above result was obtained by the first named author
\cite{Collins2002} under the assumption $n\geq d$. As we shall
see, this assumption is not necessary.

For $n\geq d$ the formula \eqref{eq:wg} takes the simpler form
\begin{equation}
\label{eq:wg2} \Wg =\frac{1}{(n!)^2} \sum_{\lambda \vdash n}
\frac{\chi^{\lambda}(e)^2}{s_{\lambda ,d}(1)}\ \chi^{\lambda},
\end{equation}
with no restrictions on the length of $\lambda$. The right-hand
side is a rational function of $d$ and hence we may consider it
for any $d\in\C$. However, the polynomial $d\mapsto
s_{\lambda,d}(1)$ has zeros in integer points
$-l(\lambda),-l(\lambda)+1,\dots, l(\lambda)-1,l(\lambda)$ and
hence the right-hand side of \eqref{eq:wg2} has poles in points
$-n,-n+1,\dots,n-1,n$ and therefore is not well-defined on the
whole $\C$.

Nevertheless, even for the case $d<n$, let us plug this incorrect
value \eqref{eq:wg2} into \eqref{bid}. In this way the right hand
side of \eqref{bid} becomes a rational function in $d$. We claim
that for every $d\in\N$ for which the left-hand side of
\eqref{bid} makes sense (i.e.\ if
$i_1,\dots,i_n,i'_1,\dots,i'_n,j_1,\dots,j_n,j'_1,\dots,j'_n\in\{1,\dots,
d\}$), the right-hand side also makes sense (possibly after some
cancellations of poles) and is equal to the left-hand side of
\eqref{bid}. Indeed, let us view the product $\Phi(A) \Phi(B) \Wg$
as an element of $\C[\Sy{n}]$ with rational coefficients in $d$.
For the choice of $A,B\in\M{d}^{\otimes n}$ used in the proof of
Corollary \ref{cor:calka} we must have $\Phi(A), \Phi(B)\in
\C_d[\Sy{n}]$ therefore the product $\Phi(A) \Phi(B) \Wg$ is an
element of $\C_d[\Sy{n}]$ with rational coefficients in $d$. Since
\eqref{eq:wg} and \eqref{eq:wg2} regarded as elements of
$\C[\Sy{n}]$ with rational coefficients in $d$ coincide on
$\C_d[\Sy{n}]$ hence our claim holds true.

We summarize the above discussion in the following proposition.
\begin{proposition}
\label{prolongation} For fixed values of the indices $\mathbf{i},
\mathbf{j},\mathbf{i}', \mathbf{j}'$ the integral
$$ \int_{\U(d)}U_{i_{1}j_{1}} \cdots
U_{i_{n}j_{n}} \overline{U_{i'_{1}j'_{1}}}\cdots
\overline{U_{i'_{n}j'_{n}}}\ dU $$ is a rational function of $d$.

Furthermore, the equation \eqref{bid} remains true (possibly after
some cancellations of poles) if we replace the correct value
\eqref{eq:wg} of Weingarten function by \eqref{eq:wg2}.
\end{proposition}

\begin{example}
Corollary \ref{cor:calka} implies that for $d \geq 2$
\begin{multline*}
\int_{\U(d)} |U_{11}|^2 \ dU = \int_{\U(d)} U_{11} U_{11}
\overline{U_{11}} \overline{ U\gwia_{11}} \ dU=\\  2 \Wg \left(
\genfrac{}{}{0pt}{1}{1}{1}\ \genfrac{}{}{0pt}{1}{2}{2} \right)+ 2
\Wg \left( \genfrac{}{}{0pt}{1}{1}{2}\ \genfrac{}{}{0pt}{1}{2}{1}
\right)=  2 \frac{1}{d^2-1} + 2 \frac{-1}{d (d^2-1)},
\end{multline*}
where the values of the Weingarten function were computed by
\eqref{eq:wg2} and where $\big(
\genfrac{}{}{0pt}{1}{1}{\sigma(1)}\cdots
\genfrac{}{}{0pt}{1}{n}{\sigma(n)}\big)$ denotes the permutation
$\sigma$. The right-hand side appears to make no sense for $d=1$,
nevertheless after algebraic simplifications we obtain
$$ \int_{\U(d)} |U_{11}|^2 \ dU = \frac{2}{d (d+1)} $$
which is a correct value for all $d\geq 1$.
\end{example}

\subsection{Asymptotics of the Weingarten function}

In this section we compute the first order asymptotic of the
Weingarten function for large values of $d$.

Consider the algebra $\C [\Sy{n}][[d^{-1}]]$ of functions on
$\Sy{n}$ valued in formal power series in $d^{-1}$ and the vector
space
$$\mathcal{A}=\Vect \left\{ \alpha\delta_{\sigma}:
\alpha=O({d^{- |\sigma|}}) \text{ and } \alpha d^{|\sigma|} \text{
is a power series in } {d^{-2}} \right\},$$ where $|\sigma |$
denotes the minimal number of factors necessary to write $\sigma$
as a product of transpositions. By the triangle inequality
$|\sigma_1|+|\sigma_2|\geq |\sigma_1 \sigma_2|$ and the parity
property $(-1)^{|\sigma_1|} (-1)^{|\sigma_2|}=(-1)^{|\sigma_1
\sigma_2|}$, $\A$ turns out to be a unital subalgebra of $\C
[\Sy{n}][[d^{-1}]]$.

It is easy to check that $d^{-n} \Phi (\Id)\in\A$. Since $d^{-n}
\Phi(\Id)= \delta_e + O(d^{-1})$ therefore its inverse
$d^n\Wg=\sum_i \big(1-d^{-n} \Phi(\Id) \big)^i$ makes sense as a
formal power series in $d^{-1}$. The following proposition follows
immediately.

\begin{proposition}
$d^n\Wg\in\mathcal{A}$. Equivalently, for any $\sigma\in\Sy{n}$,
$\Wg (\sigma)=O(d^{-n-|\sigma |})$.
\end{proposition}

In order to find a more precise asymptotic expansion we consider
the two-sided ideal $I$ in $\A$ generated by $d^{-2}\delta_e$. It
is easy to check that the quotient algebra $\A/I$ regarded as a
vector space is spanned by vectors $d^{-|\sigma|} \delta_\sigma$.
The products of these elements are given by
$$ (d^{-|\sigma|} \delta_\sigma) (d^{-|\rho|} \delta_\rho) \cong
\begin{cases}  d^{-|\sigma \rho|} \delta_{\sigma \rho} & \text{if
} |\sigma \rho|=|\sigma|+|\rho|, \\ 0  & \text{if } |\sigma
\rho|<|\sigma|+|\rho|. \end{cases} $$

Biane \cite{Biane1997crossings} considered an algebra which as a
vector space is equal to $\C[\Sy{n}]$ with the multiplication
$$ \delta_\sigma \star \delta_{\rho}=\begin{cases}  \delta_{\sigma \rho} & \text{if
} |\sigma \rho|=|\sigma|+|\rho|, \\ 0  & \text{if } |\sigma
\rho|<|\sigma|+|\rho|. \end{cases} $$ One can easily see now that
$d^{-|\rho|} \delta_{\rho} \mapsto \delta_{\rho}$ provides an
isomorphism of $\A/I$ and Biane algebra. Under this isomorphism
$d^{-n} \Phi(\Id)$ is mapped into $\zeta=\sum_{\sigma\in\Sy{n}}
\delta_{\sigma}$. The inverse of $\zeta$ in Biane algebra is
called M\"obius function and is given explicitly by
$$ \Moeb(\sigma)= \prod_{1\leq i\leq k} c_{|C_i|-1}\ (-1)^{|C_i|-1}, $$
where $\sigma$ is a permutation with a cycle decomposition
$\sigma=C_1 \cdots C_k$ and
\begin{equation} \label{eq:catalan} c_n=\frac{(2n)!}{n! (n+1)!} \end{equation}
is the Catalan number.

\begin{corollary}
$d^{n+|\sigma |}\Wg (\sigma)=\Moeb (\sigma )+O(d^{-2})$
\end{corollary}

\section{Integration over orthogonal groups}
\label{sec:orthogonal}

\subsection{Schur--Weyl duality for orthogonal groups}

\subsubsection{Brauer algebras}
We consider the group of orthogonal matrices
$$\Ort(d)=\{M\in \GL(d),M^{-1}=M^t=M^*\}.$$
Its invariant theory has first been studied by R.~Brauer
\cite{BrauerOnalgebras} who introduced a family of algebras,
nowadays called Brauer algebras. These algebras have been at the
center of many investigations (see
\cite{BirmanWenzl,GroodBraueralgebras} and the references
therein). Some actions of these algebras lead to an analogue of
the Schur--Weyl duality in the case of the orthogonal group and
symplectic groups and for this reason they are very useful for our
purposes.

Consider $2n$ vertices arranged in two rows: the upper one with
$n$ vertices denoted by $U_1,\dots,U_n$ and the bottom row with
$n$ vertices denoted by $B_1,\dots,B_n$.

\newcommand{\wys}{1.5}

\begin{figure}[h]%[bt]
\psset{unit=1cm}   $p=$
\begin{pspicture}[](0,-\wys)(11,\wys)
\psset{arrowscale=1.5}
%\pspolygon[linestyle=dotted](0,-1)(11,0)(11,2)(0,2)
\cnodeput*(1,-\wys){q1}{$B_1$} \cnodeput*(2,-\wys){q2}{$B_2$}
\cnodeput*(3,-\wys){q3}{$B_3$} \cnodeput*(4,-\wys){q4}{$B_4$}
\cnodeput*(5,-\wys){q5}{$B_5$} \cnodeput*(6,-\wys){q6}{$B_6$}
\cnodeput*(7,-\wys){q7}{$B_7$} \cnodeput*(8,-\wys){q8}{$B_8$}
\cnodeput*(9,-\wys){q9}{$B_9$} \cnodeput*(10,-\wys){q10}{$B_{10}$}
\cnodeput*(1,\wys){r1}{$U_1$} \cnodeput*(2,\wys){r2}{$U_2$}
\cnodeput*(3,\wys){r3}{$U_3$} \cnodeput*(4,\wys){r4}{$U_4$}
\cnodeput*(5,\wys){r5}{$U_5$} \cnodeput*(6,\wys){r6}{$U_6$}
\cnodeput*(7,\wys){r7}{$U_7$} \cnodeput*(8,\wys){r8}{$U_8$}
\cnodeput*(9,\wys){r9}{$U_9$} \cnodeput*(10,\wys){r10}{$U_{10}$}
%
%\cnode(1,-1){2pt}{q1} \cnode(2,-1){2pt}{q2} \cnode(3,-1){2pt}{q3}
%\cnode(4,-1){2pt}{q4} \cnode(5,-1){2pt}{q5} \cnode(6,-1){2pt}{q6}
%\cnode(7,-1){2pt}{q7} \cnode(8,-1){2pt}{q8} \cnode(9,-1){2pt}{q9}
%\cnode(10,-1){2pt}{q10} \cnode(1,1){2pt}{r1} \cnode(2,1){2pt}{r2}
%\cnode(3,1){2pt}{r3} \cnode(4,1){2pt}{r4} \cnode(5,1){2pt}{r5}
%\cnode(6,1){2pt}{r6} \cnode(7,1){2pt}{r7} \cnode(8,1){2pt}{r8}
%\cnode(9,1){2pt}{r9} \cnode(10,1){2pt}{r10}
\nccurve[border=2pt,bordercolor=white,angleA=90,angleB=-90]{q7}{r5}
\nccurve[border=2pt,bordercolor=white,angleA=90,angleB=-90]{q8}{r8}
\ncarc[border=2pt,bordercolor=white,arcangle=90]{q1}{q3}
\ncarc[border=2pt,bordercolor=white,arcangle=90]{q2}{q5}
\ncarc[border=2pt,bordercolor=white,arcangle=90]{q4}{q6}
\ncarc[border=2pt,bordercolor=white,arcangle=90]{q9}{q10}
\ncarc[border=2pt,bordercolor=white,arcangle=-90]{r1}{r3}
\ncarc[border=2pt,bordercolor=white,arcangle=-90]{r2}{r4}
\ncarc[border=2pt,bordercolor=white,arcangle=-90]{r6}{r10}
\ncarc[border=2pt,bordercolor=white,arcangle=-90]{r7}{r9}
\end{pspicture}
\caption{Example of an element of $P_{20}$} \label{fig:chip1}
%\caption{Example of a $10$-diagram.} \label{fig:chip1}
\end{figure}
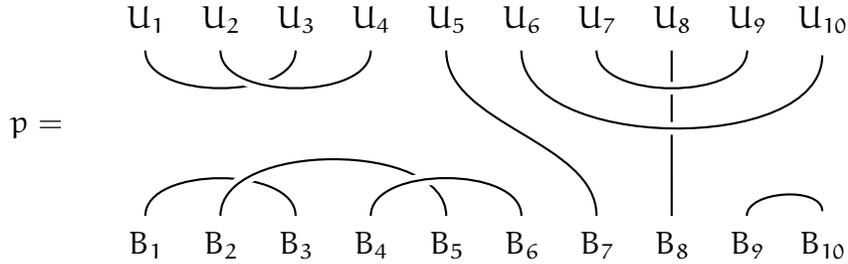

We regard $\Sy{2n}$ as a group of permutations of the set of
vertices and denote by $P_{2n}$ the set of all pairings of this
set. An example of such a pairing is presented on Figure
\ref{fig:chip1}. We can view $P_{2n}$ as a set of permutations
$\sigma\in\Sy{2n}$ such that $\sigma^2=e$ and $\sigma$ has no
fixpoints. We will consider the action $\rho_{\Sy{2n}}$ of
$\Sy{2n}$ on $P_{2n}$ by conjugation under the embedding
$P_{2n}\subset \Sy{2n}$ described above. By $\C [P_{2n}]$ we
denote the linear space spanned by $P_{2n}$. We equip this linear
space with a bilinear symmetric form $\langle \cdot, \cdot
\rangle$ by requirement that elements of $P_{2n}$ form an
orthonormal basis.
%the following line is added by B.
The embedding $P_{2n}\subset \Sy{2n}$ extends linearly to the
inclusion of $\Sy{2n}$--modules $\C [P_{2n}]\subset \C [\Sy{2n}]$
and the scalar product can be described as
$$\langle a,b \rangle=\frac{\chi^{\text{reg}}(ab^{*})}{\chi^{\text{reg}}(e)},$$
where $\chi^{\text{reg}}$ denotes the character of the left
regular representation.

%\begin{figure}[btbid]
%\psset{unit=1cm}   $Id=$
%\begin{pspicture}[](0,-\wys)(11,\wys)
%\psset{arrowscale=1.5}
%%\pspolygon[linestyle=dotted](0,-1)(11,0)(11,2)(0,2)
%\cnodeput*(1,-\wys){q1}{$B_1$} \cnodeput*(2,-\wys){q2}{$B_2$}
%\cnodeput*(3,-\wys){q3}{$B_3$} \cnodeput*(4,-\wys){q4}{$B_4$}
%\cnodeput*(5,-\wys){q5}{$B_5$} \cnodeput*(6,-\wys){q6}{$B_6$}
%\cnodeput*(7,-\wys){q7}{$B_7$} \cnodeput*(8,-\wys){q8}{$B_8$}
%\cnodeput*(9,-\wys){q9}{$B_9$} \cnodeput*(10,-\wys){q10}{$B_{10}$}
%\cnodeput*(1,\wys){r1}{$U_1$} \cnodeput*(2,\wys){r2}{$U_2$}
%\cnodeput*(3,\wys){r3}{$U_3$} \cnodeput*(4,\wys){r4}{$U_4$}
%\cnodeput*(5,\wys){r5}{$U_5$} \cnodeput*(6,\wys){r6}{$U_6$}
%\cnodeput*(7,\wys){r7}{$U_7$} \cnodeput*(8,\wys){r8}{$U_8$}
%\cnodeput*(9,\wys){r9}{$U_9$} \cnodeput*(10,\wys){r10}{$U_{10}$}

%\nccurve[border=2pt,bordercolor=white,angleA=90,angleB=-90]{q1}{r1}
%\nccurve[border=2pt,bordercolor=white,angleA=90,angleB=-90]{q2}{r2}
%\nccurve[border=2pt,bordercolor=white,angleA=90,angleB=-90]{q3}{r3}
%\nccurve[border=2pt,bordercolor=white,angleA=90,angleB=-90]{q4}{r4}
%\nccurve[border=2pt,bordercolor=white,angleA=90,angleB=-90]{q5}{r5}
%\nccurve[border=2pt,bordercolor=white,angleA=90,angleB=-90]{q6}{r6}
%\nccurve[border=2pt,bordercolor=white,angleA=90,angleB=-90]{q7}{r7}
%\nccurve[border=2pt,bordercolor=white,angleA=90,angleB=-90]{q8}{r8}
%\nccurve[border=2pt,bordercolor=white,angleA=90,angleB=-90]{q9}{r9}
%\nccurve[border=2pt,bordercolor=white,angleA=90,angleB=-90]{q10}{r10}

%\end{pspicture}
%\caption{Identity in Brauer algebra.} \label{fig:identity}
%\end{figure}

The Brauer algebra $B(d,n)$ regarded as a vector space is
isomorphic to $\C(P_{2n})$. The multiplication in the algebra
$B(d,n)$ depends on the parameter $d$, but in this article we will
not use the multiplicative structure of the Brauer algebra.

\subsubsection{Canonical representation of the Brauer algebra}
\label{subsubsec:representation}
 By $\langle \cdot , \cdot
\rangle$ we denote the canonical bilinear symmetric forms on
$\C^d$ and on $(\C^d)^{\otimes n}$. The canonical representation
$\rho_{B}$ of the Brauer algebra $B(d,n)$ on $(\C^d)^{\otimes n}$
is defined as follows: in order to compute $\langle u_1\otimes
\cdots \otimes u_n , \rho_{B}(p) [b_1\otimes \cdots \otimes b_n]
\rangle $, where $p\in P_{2n}$ and
$u_1,\dots,u_n,b_1,\dots,b_n\in\C^d$ we assign to the upper
vertices of $p$ vectors $u_1,\dots,u_n$ and to bottom vertices
vectors $b_1,\dots,b_n$. The value of $\langle u_1\otimes \cdots
\otimes u_n , \rho_{B}(p) [b_1\otimes \cdots \otimes b_n] \rangle
$ is defined to be a product of the scalar products of vectors
assigned to vertices joined by the same line. For example, for the
diagram $p$ from Figure \ref{fig:chip1} we
obtain: % (cf Figure \ref{fig:chip12}):
\begin{multline}
\label{eq:exampleofrepresentation} \big\langle u_1 \otimes \cdots
\otimes u_{10} , \rho_{{B}}(p) [b_1 \otimes \cdots \otimes b_{10}]
\big\rangle= \langle u_1,u_3 \rangle  \langle  u_2,u_4 \rangle
\times \\ \langle  u_5 ,b_7 \rangle \langle u_6 ,u_{10} \rangle
\langle  u_7 ,u_9 \rangle \langle u_8,b_8 \rangle \langle b_1,b_3
\rangle \langle b_2,b_5 \rangle \langle b_4,b_6    \rangle \langle
b_9 ,b_{10}  \rangle.
$$
\end{multline}

In the above construction we used implicitly the isomorphism of
vector spaces
\begin{equation}
\label{eq:izomorfizm} \End(\C^d)^{\otimes n}
=\bigotimes_{i\in\{U_1,\dots,U_n,B_1,\dots,B_n\}} \C^d.
\end{equation}
We will consider the action of $\Sy{2n}$ on $\End(\C^d)^{\otimes
n}$ by permutation of factors on the right-hand side of
\eqref{eq:izomorfizm}.

%WARNING: REDUNDANCY OF N-DIAGRAMS AND PAIR PARTITIONS!!!!

We consider the representation $\repo$ of $\Ort(d)$ on
$(\C^d)^{\otimes n}$, where
$$\repo (O):v_1\otimes\cdots\otimes v_n\mapsto
O( v_{1})\otimes\cdots\otimes O(v_{n})$$ is the diagonal action.
\begin{theorem}[Schur--Weyl duality for orthogonal groups
\cite{BrauerOnalgebras,Wenzl}] \label{theo:sworthogonal} The
commutant of $\repo(\Ort(d))$ is equal to $\rho_{\B} \big(
\C[P_{2n}] \big)$. Furthermore if $d\geq n$ then $\rho_{{\B}}$ is
injective.
\end{theorem}

\subsection{Integration formula}

\subsubsection{}

For $A\in \End(\C^d)^{\otimes n}$ we define
$$ \E(A)= \int_{\Ort(d)} O^{\otimes n} A (O^{t})^{\otimes n} \ dO. $$

\begin{proposition}
\label{prop:conditional2} $\E$ is a conditional expectation of \/
$\End (\C^d)^{\otimes n}$ into \linebreak $\rho_{\B}\big(\C
[P_{2n}]\big)$, in particular it satisfies $\E^2=\E$. We regard
$\End (\C^d)^{\otimes n}$ as a Euclidean space with a scalar
product $\langle A, B\rangle= \Tr A B^*$. Then $\E$ is an
orthogonal projection onto $\rho_B\big( \C [P_{2n}] \big).$  It is
compatible with the trace in the sense that
\begin{equation*}
\Tr\circ \E =\Tr.
\end{equation*}
\end{proposition}

\begin{proof}
Proof is analogous to the proof of Proposition
\ref{prop:conditional} but instead of Theorem \ref{theo:swunitary}
we use  Theorem \ref{theo:sworthogonal}.
\end{proof}

For $A\in \End(\C^d)^{\otimes n}$ we set
\begin{equation}
\label{eq:phi} \Phi(A) = \sum_{p \in P_{2n}}  p \Tr (\rho_B(p)^t
A) \in \C [P_{2n}]
\end{equation}

By the representation $\rho_B$ every element of $\C(P_{2n})$ can
be viewed as an element of $\End(\C^d)^{\otimes n}$ and therefore
we can consider the linear map
$$ \widetilde{\Phi} =\Phi\circ\rho_{\B}: \C(P_{2n}) \rightarrow \C(P_{2n}). $$
The matrix of the operator $\widetilde{\Phi}$ coincides with the
Gramm matrix of the set of vectors $\rho_{\B}(p)\in
\End(\C^d)^{\otimes n}$ indexed by $p\in P_{2n}$. We denote by
$\Wg$ the inverse of $\widetilde{\Phi}$. We postpone the problem
if this inverse exists to Proposition \ref{prop:itsok}.

We denote by $\Pi_{p_1,p_2}$ the partition induced by the action
of the group generated by $p_1,p_2$.
\begin{proposition}
\label{prop:morphisms} $\rho_{\B}$, $\E$, $\Phi$ are morphisms of
$\Sy{2n}$-spaces. %the next line is added in new version
As a consequence, $\langle p_1,\Wg p_2\rangle$ depends only on the
conjugacy class of $p_1 p_2$.
\end{proposition}

\begin{proof}
The proof of this proposition is straightforward.
\end{proof}

By a change of labels we can view $P_{2 n}$ as the set of pairings
of the set $\{1,\dots,2n\}$. We do not care about the choice of
the way in which labels $\{U_1,\dots,U_n,B_1,\dots,B_n\}$ are
replaced by $\{1,\dots,2n\}$. For a tuple of indices
$\mathbf{i}=(i_1,\dots,i_{2n})$, where
$i_1,\dots,i_{2n}\in\{1,\dots,d\}$ and a pairing $p\in P_{2n}$ we
set $\delta^{p}_{\mathbf{i}}=1$ if for each pair
$a,b\in\{1,\dots,2n\}$ connected by $p$ we have $i_a=i_b$;
otherwise we set $\delta^{p}_{\mathbf{i}}=0$.

\begin{corollary}
\label{cor:weingarten} The following formulas hold true:
\begin{equation} \label{eq:strangename0} \E =\rho_{\B} \circ \Wg \circ
\Phi,
\end{equation}
\begin{equation}
\label{eq:strangename} \Tr A \E(B)= \sum_{p_1,p_2\in P_{2n}} \Tr
\big( A \rho_{\B}(p_1) \big) \Tr \big(  \rho_{\B}(p_2)^t B\big)
\langle p_1, \Wg p_2 \rangle.
\end{equation}

For every choice of $u_1,\dots,u_{2n},v_1,\dots,v_{2n}$ we have
\begin{multline}
\label{eq:integralofapolynomial} \int_{O(d)} \langle u_1, O v_1
\rangle \cdots \langle u_{2n}, O v_{2n} \rangle \ dO =\\
\sum_{p_1,p_2\in P_{2n}} \langle u_1 \otimes \cdots \otimes u_n,
\rho_B(p_1)\
u_{n+1} \otimes \cdots\otimes u_{2n} \rangle \times \\
\langle v_1 \otimes \cdots v_n, \rho_B(p_2) \ v_{n+1} \otimes
\cdots\otimes v_{2n} \rangle\ \langle p_1, \Wg p_2 \rangle.
\end{multline}
In particular, for every choice of indices
$\mathbf{i}=(i_1,\dots,i_{2n})$, $\mathbf{j}=(j_1,\dots,j_{2n})$
\begin{equation}
\label{eq:integraloverorthogonal} \int_{\Ort(d)} O_{i_1 j_1}
\cdots O_{i_{2n} j_{2n}} \ dO = \sum_{p_1,p_2\in P_{2 n}}
\delta^{p_1}_{\mathbf{i}} \delta^{p_2}_{\mathbf{j}} \langle p_1,
\Wg p_2 \rangle.
\end{equation}

The moments of an odd number of factors vanish:
\begin{equation} \label{eq:odd} \int_{\Ort(d)} O_{i_1 j_1} \cdots O_{i_{2n+1} j_{2n+1}} \
dO = 0.
\end{equation}

\end{corollary}

\begin{proof}
It is enough to take appropriate matrices in the canonical basis
to establish this result.

The map $\Ort(d):O\mapsto -O\in \Ort(d)$ preserves the Haar
measure therefore
$$\int_{\Ort(d)} O_{i_1 j_1} \cdots O_{i_{2n+1}
j_{2n+1}} \ dO = \int_{\Ort(d)} (-O_{i_1 j_1}) \cdots
(-O_{i_{2n+1} j_{2n+1}}) \ dO $$ which shows \eqref{eq:odd}.
\end{proof}

Therefore $\Wg$ appears to be of fundamental importance in the
computation of moments of the orthogonal group, and it is of
theoretical importance to give a closed formula for it. We shall
do this in the following.

\subsubsection{An abstract formula for the orthogonal Weingarten function}\label{abstract-formula}

Let $\Id\in P_{2n}$ be any fixed pairing; to have a concrete
example let us say that $\Id$ is the identity of the Brauer
algebra, i.e.~the pairing which connects the pairs of vertices
$U_i, B_i$ with each $1\leq i\leq n$.

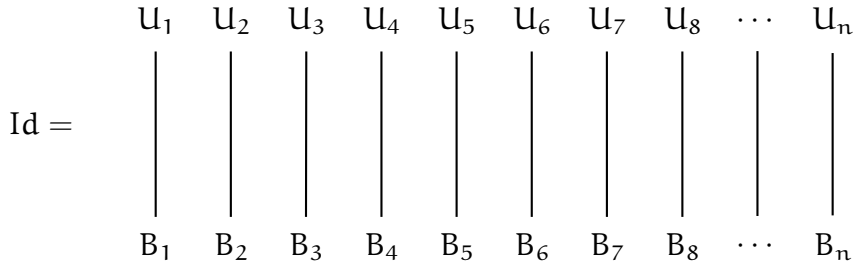
\begin{figure}[h]
\psset{unit=1cm}   $Id=$
\begin{pspicture}[](0,-\wys)(11,\wys)
\psset{arrowscale=1.5}
%\pspolygon[linestyle=dotted](0,-1)(11,0)(11,2)(0,2)
\cnodeput*(1,-\wys){q1}{$B_1$} \cnodeput*(2,-\wys){q2}{$B_2$}
\cnodeput*(3,-\wys){q3}{$B_3$} \cnodeput*(4,-\wys){q4}{$B_4$}
\cnodeput*(5,-\wys){q5}{$B_5$} \cnodeput*(6,-\wys){q6}{$B_6$}
\cnodeput*(7,-\wys){q7}{$B_7$} \cnodeput*(8,-\wys){q8}{$B_8$}
\cnodeput*(9,-\wys){q9}{$\cdots$}
\cnodeput*(10,-\wys){q10}{$B_{n}$} \cnodeput*(1,\wys){r1}{$U_1$}
\cnodeput*(2,\wys){r2}{$U_2$} \cnodeput*(3,\wys){r3}{$U_3$}
\cnodeput*(4,\wys){r4}{$U_4$} \cnodeput*(5,\wys){r5}{$U_5$}
\cnodeput*(6,\wys){r6}{$U_6$} \cnodeput*(7,\wys){r7}{$U_7$}
\cnodeput*(8,\wys){r8}{$U_8$} \cnodeput*(9,\wys){r9}{$\cdots$}
\cnodeput*(10,\wys){r10}{$U_{n}$}

\nccurve[border=2pt,bordercolor=white,angleA=90,angleB=-90]{q1}{r1}
\nccurve[border=2pt,bordercolor=white,angleA=90,angleB=-90]{q2}{r2}
\nccurve[border=2pt,bordercolor=white,angleA=90,angleB=-90]{q3}{r3}
\nccurve[border=2pt,bordercolor=white,angleA=90,angleB=-90]{q4}{r4}
\nccurve[border=2pt,bordercolor=white,angleA=90,angleB=-90]{q5}{r5}
\nccurve[border=2pt,bordercolor=white,angleA=90,angleB=-90]{q6}{r6}
\nccurve[border=2pt,bordercolor=white,angleA=90,angleB=-90]{q7}{r7}
\nccurve[border=2pt,bordercolor=white,angleA=90,angleB=-90]{q8}{r8}
\nccurve[border=2pt,bordercolor=white,angleA=90,angleB=-90]{q9}{r9}
\nccurve[border=2pt,bordercolor=white,angleA=90,angleB=-90]{q10}{r10}

\end{pspicture}
\caption{Identity in Brauer algebra.} \label{fig:identity}
\end{figure}

From now on we fix an inclusion of the hyperoctahedral group $O_n$
into $\Sy{2n}$ by considering $O_n$ as the global stabilizer of
$\Id$ under the action of $\Sy{2n}$.

We equip the set $P_{2n}$ of pairings with a metric $l$ by setting
$$ l(p_1,p_2) = \frac{|p_1 p_2|}{2}, $$
where pairings $p_1,p_2$ are regarded on the right-hand side as
elements of $\Sy{2n}$.

\begin{lemma}
\label{lem:iamsad}

If $p_1,p_2\in P_{2n}$ then
\begin{equation} \label{eq:slad}
\Tr \rho_B(p_1) \rho_B(p_2)^t =  d^{n-l(p_1,p_2)}. \end{equation}
Furthermore, $l(p_1,p_2)$ is an integer number.

Each right class $\pi O_n$ of $\Sy{2n}/O_n$ is uniquely determined
by its action $\pi( \Id)$ on the identity diagram hence the right
classes $\Sy{2n}/O_n$ are in one-to-one correspondence with the
elements of $P_{2 n}$.

We set $|\pi O_n|=\min_{\sigma\in\pi O_n} |\sigma|$. Then
\begin{equation}
\label{eq:slad2} \big\langle \widetilde{\Phi}\big(\pi (\Id) \big),
\Id \big\rangle = d^{n-|\pi O_n|}.
\end{equation}

Let a left and right class $O_n \rho O_n$ be fixed. The value of
$|\pi O_q|$ does not depend on the choice of $\pi\in O_n \rho O_n$
therefore the definition $|O_n \rho O_n|=|\rho O_n|$ makes sense.
\end{lemma}
\begin{proof}
Let $e_1,\dots,e_d$ be the orthogonal basis of $\C^d$; then
\begin{multline*} \Tr \rho_B(p_1) \rho_B(p_2)^t
=\\ \sum_{1\leq i_1,\dots,i_n,j_1,\dots,j_n\leq d} \langle
e_{j_1}\otimes \cdots \otimes e_{j_1} , \rho_B(p_1)
(e_{i_1}\otimes \cdots \otimes e_{i_q}) \rangle\times \\ \langle
e_{j_1}\otimes \cdots \otimes e_{j_1} , \rho_B(p_2)
(e_{i_1}\otimes \cdots \otimes e_{i_q}) \rangle.
\end{multline*}
To every upper vertex $U_k$ (respectively, bottom vertex $B_k$) we
assign the appropriate index $i_k$ (respectively, $j_k$). From the
very definition of $\rho_B$, the right-hand side is equal to $1$
if the indices corresponding to each pair of vertices connected by
$p_1$ or $p_2$ are equal; otherwise the right-hand side is equal
to $0$. It follows that
$$\Tr \rho_B(p_1) \rho_B(p_2)^t= d^{\text{number of connected
components of the graph depicting $p_1$ and $p_2$}}. $$ We observe
that each connected component of the graph depicting $p_1$ and
$p_2$ corresponds to a pair of orbits of the permutation $p_1
p_2$. The number of orbits of $p_1 p_2$ is equal to $2n- |p_1
p_2|$ which finishes the proof of the first part.

The above considerations imply that
$$\big\langle \widetilde{\Phi}\big(\pi (\Id) \big), \Id
\big\rangle = d^{n-\frac{1}{2} \cdot |\pi \Id \pi^{-1} \Id|}.
$$

Let $\sigma\in \pi O_n$. Since
$$|\pi \Id \pi^{-1} \Id|= |\sigma\Id\sigma^{-1} \Id|\leq |\sigma|+|\Id \sigma^{-1} \Id|=2|\sigma| $$
therefore
$$ |\pi \Id \pi^{-1} \Id| \leq 2 \ |\pi O_q|.$$

We can decompose the set of vertices
$\{U_1,\dots,U_n,B_1,\dots,B_n\}$ into two classes in such a way
that the graph depicting pairings $\pi(\Id)$ and $\Id$ is
bipartite, or---in other words---each of the pairings
$\pi(\Id),\Id$ regarded a permutation maps these two classes into
each other. We leave it to the reader to check that there exists a
unique permutation $\sigma\in \pi O_q$ which is equal to identity
on the first of these classes. It follows
$$|\pi \Id \pi^{-1} \Id| =  2 \ |\sigma| $$
which shows that
$$|\pi \Id \pi^{-1} \Id|  \geq  2 \ |\pi O_n|. $$

Let $\sigma\in O_n$. Then
$$ |\pi \Id \pi^{-1} \Id|=|\pi \Id \pi^{-1} \sigma^{-1} \Id \sigma|=
|\sigma \pi \Id \pi^{-1} \sigma^{-1} \Id|$$ therefore $|\pi
O_n|=|\sigma \pi O_n|$ finishes the proof.

\end{proof}

% KTORE KLASY SA LEWE, A KTORE PRAWE?
%
%
%
%
%B: what does it mean ? (really, I should learn Polish some day ;-)
%
%
%

\begin{lemma}\label{rsk} The sum of dimensions of representations
of\/ $\Sy{2n}$ of shape $2y_1\geq 2y_2\geq \ldots$, where
$y_1+y_2+\ldots =n$ equals the cardinality of $P_{2n}$.
\end{lemma}

\begin{proof}
The Robinson--Schensted--Knuth algorithm provides a bijection
between permutations and pairs $(P,Q)$ of standard Young tableaux
of the same shape. Furthermore if $\sigma\mapsto (P,Q)$ then
$\sigma^{-1}\mapsto (Q,P)$; it follows that the RSK algorithm is a
bijection between involutions $\sigma=\sigma^{-1}$ and standard
Young tableaux.
%
%\textbf{OK, now I have no idea how to see this condition on the
%shape of the Young diagram. Sorry.}

%
%Do you mean that you don't understand what is above ?
%I rewrite what is below. B.

%

It is easy to show that for any idempotent without fixed point,
the RSK algorithm which gives a pair of tableaux $(P,Q)$ of same
shape satisfies the additional property that $P=Q$. Furthermore,
implementing the reverse of RSK algorithm (see
\cite{FultonYoungtableaux}) shows that the tableaux must have the
shape prescribed in the Lemma, and that any such tableau gives
rise to an idempotent without fixed point.
\end{proof}

\begin{proposition}\label{p2nsplit}
The space $\C(P_{2n})$ splits under the action of\/ $\Sy{2n}$ as a
direct sum of representations associated to Young diagrams of the
shape $2y_1\geq 2y_2\geq \ldots$, where $y_1+\ldots +y_q=n$, hence
the action is multiplicity--free.
\end{proposition}

\begin{proof}
Following Fulton \cite{FultonYoungtableaux}, let us consider a
diagram of shape $2y_1\geq 2y_2\geq \ldots$ and consider its row
numbering Young tableau. Let $C$ be the column invariant subgroup
of $\Sy{2n}$ and $L$ the line invariant subgroup; both these
groups are isomorphic to a product of symmetric group. We consider
the projection operator $p_C$ associated to the trivial
representation of $C$ and the projection operator $p_L$ associated
to the alternate representation of $L$. One can see geometrically
that these two operators commute and that the partition
$(1,2)(3,4),\ldots ,(2n-1,2n)$ is not in the kernel of $p_C\circ
p_L$.

The dimension argument of Lemma \label{rsk} concludes the proof
and shows uniqueness of the occurrence of any representation of
shape $2y_1\geq 2y_2\geq \ldots$.
\end{proof}

\begin{proposition}
The eigenspaces of $\widetilde{\Phi}$ are indexed by Young
diagrams $\lambda$ with the shape $2l_1\geq 2l_2\geq\ldots$.  The
corresponding eigenvalue is given by
\begin{equation}\label{dl}
\eigen=\frac{\sum_{\pi\in O_n\backslash\Sy{2n}/O_n} d^{n-|\pi| }
\sum_{\sigma\in\pi}\chi^{\lambda}(\sigma)}{\sum_{\sigma\in
O_n}\chi^{\lambda}(\sigma)}
\end{equation}
and the corresponding eigenspace is equal to the image of
$\rho_{\Sy{2n}}(p^{\lambda})$.
\end{proposition}

\begin{proof}
$\widetilde{\Phi}$ is a morphism of $\Sy{2n}$--spaces by
Proposition \ref{prop:morphisms}, hence Proposition \ref{p2nsplit}
gives the classification of the eigenspaces of $\widetilde{\Phi}$.
Let $\lambda$ be as in Proposition \ref{p2nsplit}; then the
element $\rho_{\Sy{2n}}(p_{\lambda})(\Id)$ is non-zero and belongs
to an irreducible submodule of $\C (P_{2n})$ thus it satisfies
$$\widetilde{\Phi}\big(\rho_{\Sy{2n}}(p^{\lambda})(\Id)\big)=
\eigen \rho_{\Sy{2n}}(p^{\lambda})(\Id).$$ We have therefore by
bilinearity
\begin{equation}
\label{eq:lyon}
\langle\widetilde{\Phi}\rho_{\Sy{2n}}(p^{\lambda})(\Id),
\Id\rangle= \eigen \langle \rho_{\Sy{2n}}(p^{\lambda})(\Id),
\Id\rangle=\eigen \sum_{\sigma\in O_n } p^{\lambda}(\sigma).
\end{equation}
Lemma \ref{lem:iamsad} can be used to evaluate the left--hand side
of \eqref{eq:lyon}. Since the left--hand side of \eqref{eq:lyon}
is non--zero for sufficiently big $d$, hence also the right--hand
side is non--zero and the division makes sense.

\end{proof}

\begin{theorem}
The Weingarten function is given by
\begin{equation}
\label{eq:valueofweingerten} \Wg=\sum_{\lambda} \frac{1}{\eigen}\
\rho_{\Sy{2n}} \big(p^{\lambda}\big),
\end{equation}
where the sum runs over diagrams $\lambda$ with a shape prescribed
in Proposition \ref{p2nsplit} and $\eigen$ was defined in Equation
\eqref{dl}.

In particular,
\begin{equation}
\label{eq:valueofweingerten-bis} \langle p_1,\Wg p_2\rangle =
\sum_{\lambda} \frac{1}{z_{\lambda} (2n)!}\chi^{\text{reg}}
\big\{ \rho_{\Sy{2n}} \big(p^{\lambda}\big) (p_{1})\cdot
\rho_{\Sy{2n}} \big(p^{\lambda}\big) (p_{2}) \big\}
\end{equation}
where $\rho_{\Sy{2n}}\big(p^{\lambda}\big) (p_i)$ are considered
as elements of $\C[\Sy{2n}]$, $\cdot$ is the multiplication in
 $\C[\Sy{2n}]$.
% and $\chi^{\text{reg}}$ is the character of the
%left--regular representation of the symmetric group.
\end{theorem}

\begin{proof}
The first point follows from the above discussion and for the
second it is enough to observe that $\langle p_{1},p_{2}\rangle
=\frac{1}{(2n)!} \chi^{\text{reg}}(p_{1}p_{2}^{t})$.
\end{proof}

Observe that Equation \eqref{eq:valueofweingerten-bis} is a closed
formula for $\langle p_1,\Wg p_2\rangle$ as a (rational) function
of the parameter dimension $d$, expressed in terms of the
characters of the symmetric group (though complicated when fully
expanded - in which case the expressions of $p_{\lambda}$ and
$z_{\lambda}$ should be taken in consideration- and inconvenient
to implement on a computer).

A priori, Corollary \ref{cor:weingarten} is valid only for $d\geq
n$ since in this case $\rho_{\B}$ is injective and therefore
$\widetilde{\Phi}$ is invertible; otherwise the Weingarten
function does not exist. The following result deals also with the
cases $d<n$.

\begin{proposition}
\label{prop:itsok}
Corollary \ref{cor:weingarten} remains true for
all values of $d$ and $n$ if the following definition of the
Weingarten function is used:
\begin{equation}
\label{eq:newweingarten} \Wg=\sum_{\lambda} \frac{1}{\eigen}\
\rho_{\Sy{2n}} \big(p^{\lambda}\big),
\end{equation}
where the sum is taken over all diagrams $\lambda$ with a shape
prescribed in Proposition \ref{p2nsplit} for which $\eigen\neq 0$.
%$l(\lambda )\leq d$.
\end{proposition}
\begin{proof}
Since $\E$ is an orthogonal projection, it is enough to check the
validity of \eqref{eq:strangename0} on the range of $\rho_{\B}$.
We denote by $V\subseteq \C(P_{2n})$ the span of the images of
$\rho_{\Sy{2n}}(p^{\lambda})$ for which $\eigen\neq 0$; the range
of $\rho_{\B}$ is equal to $\rho_{\B}(V)$ hence it is enough to
show that
$$ \E \circ \rho_{\B} = \rho_{\B} \circ \Wg \circ
\widetilde{\Phi}$$ holds true on $V$. The latter equality is
obvious since the inverse of  $\widetilde{\Phi}:V\rightarrow V$ is
equal to $\Wg$ given by \eqref{eq:newweingarten}.
\end{proof}

We can treat $\widetilde{\Phi_d}:\C(P_{2n})\rightarrow \C(P_{2n})$
as a matrix the entries of which are polynomials in $d$ and
therefore its inverse $\Wg_d:\C(P_{2n})\rightarrow \C(P_{2n})$
makes sense as a matrix the entries of which are rational
functions of $d\in\C$; therefore $\Wg_d$ is well--defined for all
$d\in\C$ except for a finite set; it is explicitly given by
\eqref{eq:valueofweingerten}. For fixed $A,B\in\M{d_0}^{\otimes
n}$ let us plug this (incorrect for $d_0<n$) value of $\Wg_d$ into
\eqref{eq:strangename}; the right-hand side becomes a rational
function of $d$ and we claim that after the cancellation of poles
it has a limit $d\to d_0$ which is indeed equal to the left-hand
side of \eqref{eq:strangename}. In other words, we claim that
\begin{equation} \label{eq:regularization} \E = \lim_{d\to d_0}
\rho_{\B} \circ \Wg_{d} \circ \Phi.
\end{equation}
It is indeed the case since for every $d\in\C$ the value of $\Wg_d
\big( \Phi(A) \big)$ is the same no matter if we use
\eqref{eq:valueofweingerten} or \eqref{eq:newweingarten}.

We summarize the above discussion in the following proposition.
\begin{proposition} Corollary \ref{cor:weingarten}
remains true for all values of $d$ and $n$ if the Weingarten
function is regarded as a rational function computed in
\eqref{eq:valueofweingerten}; possibly after some cancellation of
poles.
\end{proposition}

\subsection{Asymptotics of Weingarten function}

For pairings $p_1,p_2\in P_{2n}$ let $2n_1,2n_2,\ldots$ denote the
numbers of elements in the orbits of the action of $\{p_1,p_2\}$.
We define the M\"obius function
$$ \Moeb(p_1,p_2)=  \prod_{i} (-1)^{n_i-1} c_{n_i-1},$$
where $c_n$ is the Catalan number defined in \eqref{eq:catalan}.

\begin{lemma}\label{asymptexp}
For every $p\in P_{2n}$ and $|d|$ sufficiently large we have
\begin{multline*} \Wg(p) = \\
d^{-n}   \sum_{k\geq 0} \sum_{\substack{p=p_0,p_1,\dots,p_k \\
p_{i}\neq p_{i+1} \text{ for } i\in\{0,1,\dots,k-1\}}}    (-1)^k
%(\tr p_0 p_1\gwia) (\tr p_1 p_2\gwia) \cdots (\tr p_{n-1} p_n\gwia) p_n
d^{-\dist(p_0,p_1)-\cdots-\dist(p_{n-1},p_{n})} p_n.
\end{multline*}
\end{lemma}
\begin{proof}
It is enough to observe
$$d^{-n} \widetilde{\Phi}(p)= p+ \sum_{p'\neq p} d^{-l(p,p')} p'  $$
and use the power series expansion
$\frac{1}{1+x}=1-x+x^2-x^3+\cdots$ for the operator $d^{-n}
\widetilde{\Phi}$.
\end{proof}

\begin{theorem}\label{asymp-ortho}
The leading term of the Weingarten function is given by
\begin{equation}
\label{eq:leading} \langle p, \Wg p' \rangle =
d^{-n-\dist(p,p')}\Moeb(p,p') +O(d^{-n-\dist(p,p')-1}).
\end{equation}
\end{theorem}
\begin{proof}
Lemma \ref{asymptexp} implies that we need to find explicitly all
tuples of pairings $p_0,\dots,p_{k}$ such that $p_0=p$, $p_k=p'$
which fulfill $p_i\neq p_{i+1}$ for $i\in\{0,\dots,k-1\}$ and
$l(p_0,p_1)+\cdots+l(p_{k-1},p_k)=l(p_0,p_k)$.

For every such tuple the triangle inequality implies that
$\dist(p_0,p_i)+\dist(p_i,p_k)=\dist(p_0,p_k)$, or equivalently,
$|p_0 p_i|+|p_i p_k|=|p_0 p_k|=|(p_0 p_i)(p_i p_k)|$. The latter
condition implies that every orbit of ${p_0} {p_i}\in\Sy{2n}$ must
be a subset of one of the orbits of $p_0 p_k$
\cite{Biane1997crossings,Biane1998}. Therefore pairing $p_i$
cannot connect vertices which belong to different connected
components of the graph spanned by $p_0$ and $p_k$. It follows
that it is enough to consider the case if the graph spanned by
$p_0$ and $p_k$ is connected.

Suppose that the graph spanned by $p_0$ and $p_k$ is connected. It
follows that the permutation $p_0 p_k$ consists of two $n$-cycles,
we denote one of them by $\pi$. Since every orbit of $p_0 p_i$ is
a subset of one of the orbits of $p_0 p_k$ therefore it makes
sense to consider the restriction $\rho_i$ of $p_0 p_i$ to the
support of $\pi$. Observe that knowing $\rho_i$ we can reconstruct
the pairing $p_i$ by the formula
$$ p_i(s)=\begin{cases} p_0 \rho_i(s) & \text{if $\rho_i(s)$ is defined,} \\
\rho_i^{-1} p_0(s) & \text{otherwise.}
 \end{cases} $$
It follows that the solutions of the equation
$l(p_0,p_i)+l(p_i,p_k)=l(p_0,p_k)$ can be identified with the
solutions of the equation $|\rho|+|\rho^{-1} \pi|=|\pi|$.

Now one can easily see that the tuples of pairings
$p_1,\dots,p_{k-1}$ which fulfill $p_i\neq p_{i+1}$ for
$i\in\{0,\dots,k-1\}$ and
$l(p_0,p_1)+\cdots+l(p_{k-1},p_k)=l(p_0,p_k)$ are in one-to-one
correspondence with tuples of permutations
$\rho_1,\dots,\rho_{k-1}$ such that $\rho_i\neq \rho_{i+1}$ and
$|\rho_0 \rho_1^{-1}|+\cdots+|\rho_{k-1} \rho_k^{-1}|=|\rho_0
\rho_k^{-1}|$, where $\rho_0$ is the identity permutation and
$\rho_k=\pi$. The results of Biane \cite{Biane1997crossings}
finish the proof.

\end{proof}

\subsection{Cumulants}

Recall that in the work of the first named author
\cite{Collins2002} the asymptotics of cumulants of unitary
Weingarten functions have been obtained (Theorem 2.15). The
purpose of this section is to establish the counterpart of this
result for orthogonal $\Wg$ functions.

As we see by Proposition \ref{prop:morphisms}, the function $\We$
can be labelled by $\We (\lambda ,d)$ were $\lambda\vdash n$ is a
partition of the number $n$. It will be more convenient to define
in the obvious way $\We (\pi ,d)$ where $\pi$ is a partition of the
interval $[1,n]$. For partitions $\Pi,\Pi'$ of $[1,n]$
such that $\pi\leq\Pi\leq\Pi'$, it is of
fundamental importance to have a good understanding of {\it
relative cumulants \it} $C_{\pi,\Pi,\Pi'}$ of $\We$ defined
implicitly by the relation
$$\We_{\Pi'} (\pi ,d)=\sum_{\Pi\leq\Pi''\leq\Pi'}C_{\pi,\Pi,\Pi''}$$
whenever $\Pi'\geq\Pi$, with $\We_{\Pi }(\pi)=\prod_k \We
(\pi_{|V_k})$ if one denotes $\Pi=\{V_1,\ldots ,V_k\}$.

\begin{lemma}
The relative cumulant is given for $d$ large enough, by
\begin{multline*} C_{\pi,\Pi,\Pi'}=
d^{-n}   \sum_{k\geq 0} \sum_{\substack{p=p_0,p_1,\dots,p_k \\
p_{i}\neq p_{i+1} \text{ for } i\in\{0,1,\dots,k-1\}\\
\sup (\Pi,\pi,\pi_1,\ldots ,\pi_k)=\Pi'
}}    (-1)^k
d^{-\dist(p_0,p_1)-\cdots-\dist(p_{n-1},p_{n})}
\end{multline*}

%\textbf{I am scared that you do not take some signed sum...}
%B: oops, sorry. hope this works this time.
The
leading order of the series of $C_{\pi,\Pi,\Pi'}$
is therefore the number of $k$-tuples
$(\pi_1,\ldots ,\pi_k)$ of elements of $P_{2n}$ such that
$l(\pi,\pi_1)+l(\pi_1,\pi_2)+\ldots +l(\pi_k,Id)= n+l(\pi,
\Id)-2(\#\text{blocks}(\Pi ')-\# \text{blocks}(\Pi))$ together
with the requirement that the
$$\sup (\Pi,\pi,\pi_1,\ldots ,\pi_k)=\Pi'$$
\end{lemma}

%one precision added.
\begin{proof}
For the first point, it is enough to check that this equation satisfies the
moment-cumulant Equation. Asymptotics of the leading order is elementary.
For a less direct approach, see also
\cite{Collins2002}.
\end{proof}

In order to compute the leading order,
it is enough to compute the number of $k$ -tuples $(\pi_1,\ldots
,\pi_k)$ of elements of $P_{2n}$ such that
$d(\pi,\pi_1)+d(\pi_1,\pi_2)+\ldots +d(\pi_k,Id)= n+l(\pi,
\Id)-2(\# \text{blocks}(\Pi ')-\# \text{blocks}(\Pi))$ together
with the requirement that the $\sup (\pi,\pi_1,\ldots
,\pi_k)=1_n$. We call $B[\pi,k]$ this number.

Denote by $\tau_1,\ldots ,\tau_n$ the disjoint transpositions
generating the pairing $Id\in P_{2n}$, and $G$ be the subgroup of
$\Sy{2n}$ generated by these transpositions. This group has the
structure of $(\mathbb{Z}/2\mathbb{Z})^n$.

The symmetric group $\Sy{n}$ can be regarded as a subset of
$P_{2n}$ when we identify permutation $\sigma$ with a pairing
which connects the upper vertex $U_i$ with the bottom vertex
$B_{\sigma(i)}$ for all values of $1\leq i\leq n$. We says that
pairings which can be obtained by this construction are
{\it permutation-like\it}. The group $G$ acts on $P_{2n}$ by conjugations
and one checks easily that in any orbit under the action of $G$
there exist at least one permutation-like element. Moreover, two
permutation-like element in a same orbit are conjugate to each
other when regarded as elements of $\Sy{n}$. More precisely, each
orbit has $2^l$ elements, where $l$ is the number of cycles with
at least $3$ elements in $\Sy{n}$.

Fix $\pi\in P_{2n}$ and call $k$ the number of its connected
components (i.e. the number of cycles -including trivial cycles
(two-element orbits) and transpositions (four-elements orbits) of
an associated permutation-like element).

Let $\sigma\in\Sy{n}$ be one image $\pi$. Consider the number of
$k$ -tuples $(\sigma_1,\ldots ,\sigma_k)$ of permutations of
$\Sy{n}$ such that $\sigma_1\ldots\sigma_k\sigma=e$, the group
generated by $\sigma_1,\ldots ,\sigma_k$ acts transitively on
$[1,n]$ and $|\sigma|+|\sigma_1|+\ldots +|\sigma_k|=2n-2$. This
number has already been computed in
\cite{MR2001g:05006} %reference is Bousquet-Mellou and Schaeffer
and it is
\begin{equation*}
\tilde{A}[\sigma
,k]=k\frac{(nk-n-1)!}{(nk-2n+|\sigma|+2)!}\prod_{i\geq 1} \left[ i
{ki-1 \choose i}\right]^{d_i}
\end{equation*}
where $d_i$ denotes the number of cycles with $i$ elements of
$\sigma$.

\begin{proposition}
$B[\pi ,k]=2^{k-1}\tilde{A}[\sigma ,k]$.
\end{proposition}

\begin{proof}
The group $G$ acts by conjugation on $k$-tuples $(\pi_1,\ldots
,\pi_k)$ arising in the counting of $B[\pi ,k]$. Choose one
element of $G$ that turns $\pi$ into permutation like. In other
words, one can assume that $\pi$ is permutation like. Introduce
the group $G'$ generated by
$\tau_{i_{1,1}}\ldots\tau_{i_{1,l_1}},\ldots
,\tau_{i_{k,1}}\ldots\tau_{i_{k,l_k}}$ where
$\tau_{i_{j,1}},\ldots,\tau_{i_{j,l_j}}$ correspond to elements of
the $j$-th cycle of $\sigma$. This group has the structure of
$(\mathbb{Z}/2\mathbb{Z})^k$ and acts by restriction of $G$ on the
$k$-tuples $(\pi_1,\ldots ,\pi_k)$. One checks that for any
$k$-tuple $(\pi_1,\ldots ,\pi_k)$ satisfying length conditions,
there exists two and only two elements of $G'$ such that their
action turns all $k$-tuples into permutation like elements.
\end{proof}

\begin{theorem}
\label{thm:cumulants}
$C_{\pi,\Pi,\Pi'}$ is a rational fraction of order \\
$d^{-n-l(\pi, \Id)+2(\# blocks(\Pi ')-\# blocks(\Pi))}$ whose
leading term is given by $\gamma_{\pi,\Pi,\Pi'}$. Assume that
$\pi$ has $d_i$ cycles of length $i-1$. Then
\begin{equation}\label{conj}
\gamma_{\pi, \pi, 1_n}=(-1)^{|\pi |}
\frac{2^{2q-2|\pi|-1}(3q-3-|\pi|)!}{(2q)!} \prod_{i=1}^q
\left(\frac{(2i-1)!}{(i-1)!^2} \right)^{d_i}
\end{equation}
\end{theorem}

\begin{proof}
The proof is exactly the same as that of Theorem 2.15 in
\cite{Collins2002}. It is enough, in Equation (2.56), to replace
$\tilde{A}[\sigma ,k]$ by $B[\pi ,k]$
\end{proof}

\section{Integration over symplectic groups}
Let $e_{1},\dots,e_d,f_1,\dots,f_d$ be an orthonormal basis of
$\C^{2d}$. We refer to this basis as the canonical basis. Consider
the bilinear antisymmetric form $\langle\cdot,\cdot\rangle$ such
that
\begin{equation}
\label{defsympl} \langle e_i,f_j \rangle=\delta_{i,j},\qquad
\langle e_i,e_j\rangle=\langle f_i,f_j\rangle=0
\end{equation}
 The symplectic group
$\Smp(d)$ is the set of unitary matrices of $\M{2d}$ preserving
$\langle \cdot,\cdot \rangle$. Also by $\langle \cdot,\cdot
\rangle$ we denote the bilinear form on $(\C^{2d})^{\otimes n}$
given by the canonical tensor product of forms $\langle
\cdot,\cdot \rangle$ on $\C^{2d}$. This form is symmetric if $n$
is even and antisymmetric if $n$ is odd.

The Brauer algebra ${\B}(-d,n)$ admits a natural action onto the
space $(\C^{2d})^{\otimes n}$ given in the same way as in Section
\ref{subsubsec:representation} with the difference that $\langle
\cdot,\cdot \rangle$ should be understood as in Equation
\eqref{defsympl}.

The most of the results from the section \ref{sec:orthogonal}
remain true also for the symplectic case. Below we present briefly
which changes are necessary.

\begin{theorem}[Schur--Weyl duality for symplectic groups
\cite{BrauerOnalgebras,Wenzl,BirmanWenzl}]
\label{theo:sworthosymp} The commutant of $\rho_{\Smp(d)}(\Smp(d))$
is equal to $\rho_{\B} \big( \C[P_{2n}] \big)$. Furthermore if
$d\geq n$ then $\rho_{{\B}}$ is injective.
\end{theorem}

For $A\in\End(\C^{2d})^{\otimes n}$ we set
$$ \E(A)= \int_{\Smp(d)} O^{\otimes n} A (O^{t})^{\otimes n} \ dO $$
and define $\Phi(A)$ as in \eqref{eq:phi}. All results of Section
\ref{sec:orthogonal} remain true with the only difference that the
value of $d$ in all formulas should be replaced by $(-d)$.

As for the cumulants, $\gamma_{\pi,\pi,1_n}$ should be replaced by
$(-1)^{k+1}\gamma_{\pi,\pi,1_n}$ where $k$ is the number of blocks of $\pi$.

\section{Expectation of product of random matrices and free probability}

This section is rather sketchy since it follows very closely the
work of the first--named author \cite{Collins2002}.

\subsection{Asymptotic freeness for orthogonal matrices}

Let $n$ be an integer. We consider the following enumeration of
$8n$ integers: $1,\ldots 4n,\overline{1},\ldots ,\overline{4n}$.
Consider $\mathcal{T}$ the subset of $B_{8n}$ such that any
pairing links an $i$ with a $\overline{j}$. This set is isomorphic
to $\Sy{4n}$. Call $\Xi$ the element of $B_{8n}$ linking $2i-1$ to
$2i$ and $\overline{2j-1}$ to $\overline{2j}$, and $\mathcal{S}$
the subset of $B_{8n}$ such that elements link $\overline{2i-1}$
to $\overline{2i}$ and an odd (resp. even) $j$ to an odd (resp.
even) $k$.

Let $A^{(1)},\ldots ,A^{(2n)}$ be (constant) matrices in $\M{d}$.
For $\tau\in B_{4n}$, and $B$ a random matrix, define
\begin{multline}\label{def-t}
\tr (A^{(1)},\ldots ,A^{(2n)};B,\tau)= \\
d^{-\text{loops} (\Xi ,\tau)} \E\big(\sum_{k_1,\ldots
,k_{4n},k_{\overline{1}},\ldots ,k_{\overline{4n}}} \prod_{i=1}^n
B_{k_{2i-1},k_{2i}}A^{(i)}_{k_{\overline{2i-1}},k_{\overline{2i}}}
\delta_{\tau} \big)
\end{multline}
where $\delta_{\tau}=1$ if for all pair $(i,j)$ of $\tau$,
$k_i=k_j$, and $0$ else. This expression is obviously a product of
normalized traces of $\{B,B^t\}$ alternating with
$\{A^{(i)},A^{(i)t}\}$

Let $\tau\in\mathcal{T}$ and $\sigma\in\mathcal{S}$. Define
\begin{equation}
\tr (A^{(1)},\ldots ,A^{(2n)};\tau,\sigma)= d^{-\text{loops}
(\sigma ,\tau)} \E\big(\sum_{k_1,\ldots
,k_{2n},k_{\overline{1}},\ldots ,k_{\overline{2n}}} \prod_{i=1}^n
A^{(i)}_{k_{\overline{2i-1}},k_{\overline{2i}}}
\delta_{\tau}\delta_{\sigma} \big)
\end{equation}
As in Equation \eqref{def-t} this expression is obviously a
product of normalized traces of $\{A^{(i)},A^{(i)t}\}$.

Let $O$ be a random orthogonal Haar distributed matrix in $\M{d}$.
One establishes easily
\begin{multline}\label{genmoment}
\tr (A^{(1)},\ldots ,A^{(2n)};O,\tau)= \\
\sum_{\sigma\in\mathcal{S}} \tr (A^{(1)},\ldots
,A^{(2n)};\tau,\sigma)\widetilde{\Wg} (\sigma,\Xi) d^{l(\Xi
,\tau)-l(\Xi,\sigma)-l(\sigma,\tau)}
\end{multline}
where $\widetilde{\Wg}$ is the asymptotic normalized $\Wg$
function restricted on the set $\{1,\ldots ,4n\}$. From this we
obtain

\begin{lemma}
In Equation \eqref{genmoment}, assuming that
$\{A^{(i)},A^{(i)*}\}$ admits a joint limit distribution with
respect to the normalized trace $\tr$ on $\M{d}$, any term on the
right hand side has asymptotic order $\leq 0$. In case $l(\Xi
,\tau)-l(\Xi,\sigma)-l(\sigma,\tau)=0$, at least two factors of
$\tr (A^{(1)},\ldots ,A^{(2n)};\tau,\sigma)$ have to be of the
kind $\tr (A^{(i)})$. In addition, at least two of the such
indices $i$ are such that neither the pattern $"\ldots
OA^{(i)}O^*\ldots "$ nor $"\ldots O^*A^{(i)}O\ldots "$ occurs in
the cycle decomposition.
\end{lemma}

\begin{proof}
The first point is an obvious consequence on triangle inequality.
In the case $l(\Xi ,\tau)-l(\Xi,\sigma)-l(\sigma,\tau)=0$, observe
that since $l(\Xi,\sigma)\geq n$, one has to have $l(\sigma
,\tau)\leq 3n-1$. The remaining assertions are an easy adaptation
of \cite{Collins2002}, Proposition 3.3. (note that
$l(\sigma,\tau)\geq 2n$ according to the definition of
$\mathcal{T}$ and $\mathcal{S}$ and the proof follows by an easy
graphical interpretation and the description of geodesic given in
proof of \cite{Collins2002}, Theorem 3.13)
\end{proof}

From this we deduce:

\begin{theorem}
Let $O_1, O_2.\ldots$ be independent copies of orthogonal
ensembles. And $W$ a set of matrices such that the set $(W,W^t)$
admits a limit distribution. Then
$W,\{O_1,O_1^t\},\{O_2,O_2^t\}\ldots $ are asymptotically free.
This convergence holds almost surely.
\end{theorem}

\begin{proof}
Asymptotic freeness is an immediate application of definition of
freeness together with the previous Lemma and asymptotic
multiplicativity of $\Wg$ function established at Theorem 3.13

The proof of almost sure convergence is a consequence on the
computation of cumulants of $\Wg$ function in Theorem 3.16
together with an application of Chebyshev inequality and
Borel-Cantelli lemma (see \cite{Collins2002}, Theorem 3.7 for
details).
\end{proof}

\begin{remark}
We would like to draw the attention of the reader on the fact that
the situation is not as general as for the unitary case. For
example, in the unitary case, the matrix family
$(2^dE_{i,i+1}^{(d)},\{O,O^*\})\in\M{d}$ admits an asymptotic
joint law whereas this is not true in the orthogonal case. One way
of getting around this problem is to assume that matrices are
bounded. An other option is to modify the joint law assumption by
enlarging the family $W$ to $W,W^t$ as we do in the previous
Theorem. It is also possible to write down a necessary and
sufficient relation from Equation \eqref{genmoment} but to our
knowledge, there is no mathematical need for this at this
point.
\end{remark}

\subsection{Orthogonal matrix integral}

In this section we deal with orthogonal matrix integrals, and in
particular with the orthogonal Itzykson-Zuber integral. For
unitary matrix integrals many tools are available and this paper
together with \cite{Collins2002} just provide a complementary
mathematical approach. However, interestingly enough, it seems
that up to now there were no systematic tools for the study of
non-unitary (i.e.\ orthogonal, symplectic) matrix integrals. One
bright side of our approach is to provide such a tool and
therefore new formulae to theoretical physics.

\begin{theorem}\label{cv-gen}
Let $W$ be a family of matrices such that the family $W,W^t$
admits a limit joint distribution. Let $O_1, \ldots ,O_k$ be
independent Haar distributed unitary (resp.\ orthogonal or
 symplectic) matrices. Let $(P_{i,j})_{1\leq i,j \leq k}$ and $(Q_{i,j})_{1\leq
i,j \leq k}$ be two families of noncommutative polynomials in
$O_1,O_1^*,\ldots ,O_k,O_k^*$ and $W$. Let $A_d$ be the random
variable $\sum_{i=1}^k\prod_{j=1}^k\tr P_{i,j}(O,O^*,W)$ and $B_d$
the variable $\sum_{i=1}^k\prod_{j=1}^k\tr Q_{i,j}(O,O^*,W)$,
where $\tr x=\frac{1}{d} \Tr x$ for $x\in \M{d}(\C)$ denotes the
normalized trace.
\begin{itemize}
\item (i) For each $d$, the analytic function
\begin{equation*}
z\rightarrow d^{-2}\log \E \exp (z d^2A_d) =\sum_{n\geq 1}a_{d,n}
z^n
\end{equation*}
is such that for all $n$ the limit $\lim_{d\to\infty} a_{d,n}$
exists and is finite. It depends only on the limit distribution of
$W$ and on the polynomials $P_{i,j}$. \item (ii) For each $d$, the
analytic function
\begin{equation*}
z\rightarrow \frac{\E \exp (zB_d+zd^2A_d)}{\E \exp
(zd^2A_d)}=1+\sum_{n\geq 1}b_{d,n} z^n
\end{equation*}
is such that for all $n$ the limit $\lim_{d\to\infty} b_{d,n}$
exists and is finite. It depends only on the limit distribution of
$W$ and on the polynomials $P_{i,j}$ and $Q_{i,j}$.
\end{itemize}
\end{theorem}

\begin{proof}
This is a straightforward application of Theorem 3.16. See Theorem
4.1 of \cite{Collins2002} for details.
\end{proof}

As a further illustration of our results on the asymptotics of
cumulants, we state the asymptotics of
$$d^{-2}C_n(d\Tr A_dOB_dO^*)$$
This number is also known as the coefficient of the series of the
orthogonal {\it Itzykson-Zuber integral\it}. Observe that if $A_d,B_d$ are
real symmetric, the Harish-Chandra formula
applies and yields a formula for finite dimensional IZ integral
provided that the eigenvalues of $A_d$ and $B_d$ have no
multiplicity. Without these assumptions, there is no formula to our
knowledge. However, interesting results have been obtained in
\cite{1021.82011} (see also references therein)
about asymptotics of symplectic Harisch-Chandra
integrals and the two results would deserve to be compared.

The asymptotic convergence of $d^{-2}C_n(d\Tr A_dOB_dO^*)$
provided that $A_d,A_d^t,B_d,B_d^t$ admit a joint limit
distribution is already granted by Theorem \ref{cv-gen}. Let $G_n$
be the set of (not-necessarily connected) planar graphs (such that
any connected component is drawn on a distinct sphere) with $n$
edges together with the following data and conditions:
\begin{itemize}
\item[i] each face has an even number of edges, \item[ii] the
edges are labelled from 1 to $n$, \item[iii] there is a bicoloring
in white and black of the vertices such that each black vertex has
only white neighbors and vice versa.
\end{itemize}

To each such graph $g\in G_n$ we associate the permutations
$\sigma (g)$ (resp. $\tau (g)$) of $\Sy{n}$ defined by turning
clockwise (resp. counterclockwise) around the white (resp. black)
vertices and the function
\begin{equation*}\index{$\Moeb(g)$}
\Moeb(g)=\gamma_{\tau\sigma^{-1},\Pi_{\tau}\vee\Pi_{\sigma},
q+|\tau\sigma^{-1}|+2(C(\Pi_{\tau}\vee\Pi_{\sigma})-1)}.
\end{equation*}
For this definition to make sense in the orthogonal framework, we
chose an embedding of $\Sy{n}$ into $B_{2n}$ by partitioning
$[1,2n]$ into two sets $V_1$ and $V_2$ of cardinal $n$ and to a
permutation $\sigma$, we associate an element of $B_{2n}$ pairing
the $i$th element of $V_1$ to the $\sigma (i)$th element of $V_2$.

For example in the picture,
\begin{equation*}
\begin{split}
\sigma=(1 \, 13 \, 2)(3\, 5\, 4)(6 \, 7)(8\, 9 \, 10)(11 \, 12)(16 \, 17)(14 \, 15)\\
\tau=(5\, 6)(7\, 8)(10\, 11)(2\, 3\, 9)(12 \, 13)(1\, 4)(14 \, 17)(15 \, 16)\\
\tau\sigma^{-1}=(1\, 3)(5\, 9\, 7)(6\, 8\, 11\, 13\, 4)(2\, 12\, 10)(17\, 15)(14\, 16)\\
\end{split}
\end{equation*}

%ce qui suit utilise le package graphicx
\includegraphics[scale=0.5]{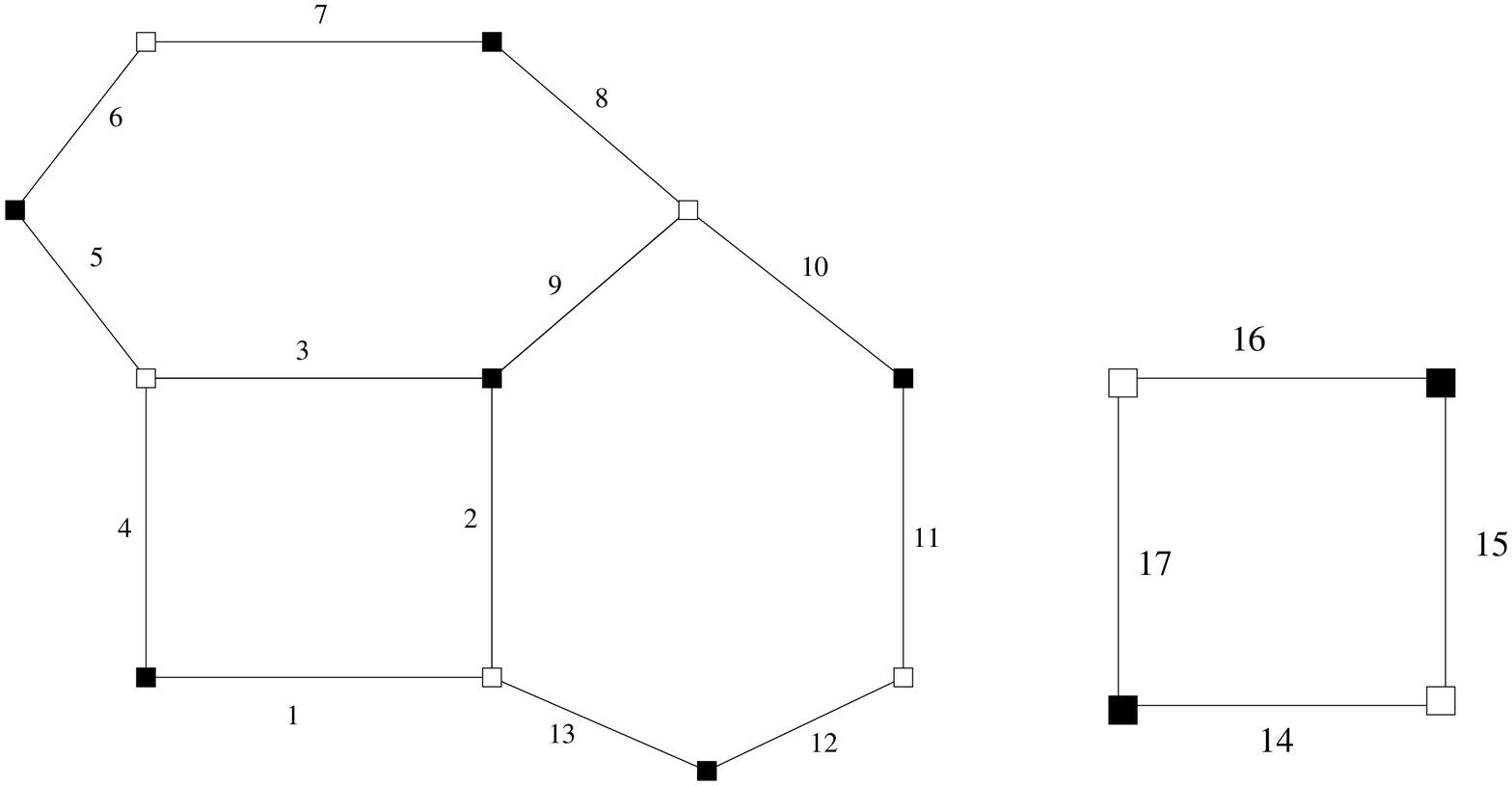}

Two graphs are said to be equivalent if there is a positive
oriented diffeomorphism of the plane transforming one to the other
and respecting the coloring of the vertices and the labelling of
the edges. We call $\sim$ this equivalence relation. For a
permutation $\sigma\in\Sy{n}$, we call $\langle X\rangle_{\sigma}$
the amount $\prod_{i=1}^k\tr X^{l_i}$ if $\sigma$ splits into
orbits containing $l_1,\ldots ,l_k$ elements.

\begin{theorem}\label{diagr}
If $X_d,Y_d,X^t_d,Y^t_d$ admit a joint limit distribution, one has
\begin{equation}
\lim_d d^{-2}C_n (d^2A)=\sum_{g\in G_q/\sim}\langle X
\rangle_{\tau(g)}\langle Y \rangle_{\sigma (g)}\Moeb(g)
\end{equation}
\end{theorem}

We omit this proof, for it is almost the same as that of Theorem
4.3 of \cite{Collins2002}. Observe that the asymptotic result only
depends on traces of polynomials in $X_d$ and traces of
polynomials in $Y_d$. Mixed patterns (involving traces of a
non-commutative polynomial in the four variables
$X_d,Y_d,X^t_d,Y^t_d$) do not occur in the limit.
%the above might deserve some justification although graphically it is rather clear
However we need a control on the joint moments. In other words,
the same diagrams appear as in the unitary case. The only
difference is that the orthogonal function $\Moeb$ is the unitary
one times $2^{\# \text{connected components} -1}$.

\begin{theorem}\label{thmfree}
Let $X_d$ be a rank one projection and assume that $(Y_d,Y_d^t)$
has a limit joint distribution whose first marginal is $\mu$.
\begin{equation}\label{final}
\lim_d d^{-1}\cdot C_n(d\Tr (X_dOY_dO^*)=(n-1)!k_n(\mu )
\end{equation}
In other words, the coefficients of $z\rightarrow d^{-1}\log \E
e^{d\Tr (X_dOY_dO^*)}$ converge pointwise to those of the
primitive of R-transform of $\mu$.
\end{theorem}

The proof goes along the same lines as Theorem 4.7 of
\cite{Collins2002}, therefore we omit it. Observe that this result
is exactly the same as for the unitary case, except that we need
an extra control on the joint moments of $X_d,Y_d,X^t_d,Y^t_d$.

\subsection{Orthogonal replaced by symplectic}

The statement when replacing orthogonal matrices by symplectic
should be replaced in the following way: if $P$ is the unitary
such that $PO^TP=O^*$, then $(X_d,X_d^t)$ (resp.
$(Y_d,Y_d^t)$)should be replaced by $(X_{2d},PX_{2d}^TP)$ (resp.
$(Y_{2d},PY_{2d}^TP)$). Theorem \ref{cv-gen} remains true. Theorem
\ref{diagr} as well (one only needs to modify accordingly the
definition of $\Moeb$).

In Theorem \ref{thmfree}, $\mu$ should be replaced
by $-\mu$.%<- I should check this detail again.
%please check this if you feel in the mood.
%
%

%
%As a concluding remark, observe that the following matrix
%integrals converge towards a non-trivial limit:

%-orthogonal
%$$d^{-3}\log \E e^{d^2\Tr (AOBO)}$$
%-symplectic
%$$d^{-3}\log \E e^{d^2\Tr (AOBO^*)}$$
%with $A,B$ only upper right quarter (2-nilpotent)
%
%
%
%
%
%
%
%These integrals are purely symplectic (or orthogonal) and do not
%have counterparts in the unitary case. Surprisingly, their
%relevant scaling seems to be $d^3$ and not $d^2$, which probably
%deserves some physical interpretation.
%
%NB: here I have not checked the mathematics... Maybe I will have to remove that...
%

\section{Examples of $\Wg$ function}

We present below the values of the Weingarten function computed
for the orthogonal group $O_d$. In order to obtain the appropriate
results for the symplectic group $\Smp_{d}$ one should replace in
the formulas $d$ by $-d$. These formulae have been obtained
directly from the definition of $\Wg$, without the help of formula
\eqref{eq:valueofweingerten-bis}. Observe that relative cumulants
that can be obtained from these value yield asyptotics predicted
by Theorem \ref{thm:cumulants}, formula \eqref{conj}.

\begin{align*} \Wg ([1])& =d^{-1},\\
\Wg ([1,1])& =\frac{d+1}{d(d-1)(d+2)},\\
\Wg ([2])& =\frac{-1}{d(d-1)(d+2)},\\
\Wg ([1,1,1])& =\frac{d^2+3d-2}{d(d-1)(d-2)(d+2)(d+4)},\\
\Wg ([2,1]) &=\frac{-1}{d(d-1)(d-2)(d+4)},\\
\Wg ([3]) &=\frac{2}{d(d-1)(d-2)(d+2)(d+4)},\\
\Wg ([4])& =\frac{-5d-6}{d(d+1)(d+2)(d+4)(d+6)(d-1)(d-2)(d-3)},\\
\Wg ([3,1])& =\frac{2d+8}{(d+1)(d+2)(d+4)(d+6)(d-1)(d-2)(d-3)},\\
\Wg ([2,2])& =\frac{d^2+5d+18}{d(d+1)(d+2)(d+4)(d+6)(d-1)(d-2)(d-3)},\\
\Wg
([2,1,1])& =\frac{-d^3-6d^2-3d+6}{d(d+1)(d+2)(d+4)(d+6)(d-1)(d-2)(d-3)},\\
\Wg ([1,1,1,1])&
=\frac{d^4+7d^3+d^2-35d-6}{d(d+1)(d+2)(d+4)(d+6)(d-1)(d-2)(d-3)}.
\end{align*}

%NB: it would be worthwhile to mention the joint works of Hikami-Brezin
%on orthogonal and symplectic IZ integral, and compare

\section{Acknowledgements}

B.C. was Allocataire Moniteur at the Ecole Normale Sup\'erieure,
Paris while a part of this work was done. He is currently a JSPS
postdoctoral fellow.

P.\'S.~was supported by State Committee for Scientific Research
(KBN) grant No.\ 2 P03A 007 23. Research of P.\'S.~was performed
during a visit in Ecole Normale Sup\'erieure (Paris) and Institute
des Hautes Etudes Scientifiques funded by European Post-Doctoral
Institute for Mathematical Sciences.

\bibliographystyle{alpha}
\bibliography{unitary-final}

\end{document}